\documentclass[aps,pra,superscriptaddress,longbibliography,twocolumn]{revtex4-1}

\usepackage{graphicx,hyperref,color}

\newcommand{\drdf}[1]{#1}

\newcommand{\figA}{
	\begin{figure*}[t]
	\includegraphics{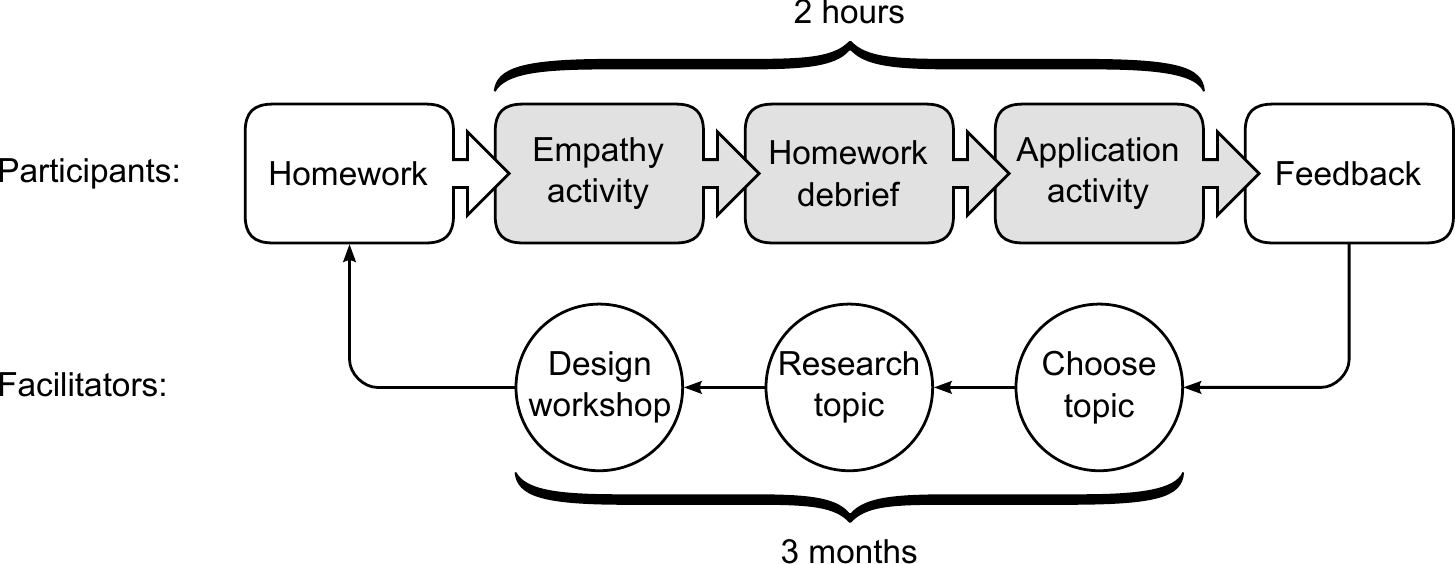}
	\caption{\label{fig:model}Workshop flowchart. Rounded square blocks represent the participant experience. White blocks represent pre- and post-workshop activities; shaded blocks correspond to activities during the workshop. Circles represent facilitation tasks.}
	\end{figure*}
}

\newcommand{\figB}{
	\begin{figure}[t]
	\includegraphics[width=\columnwidth]{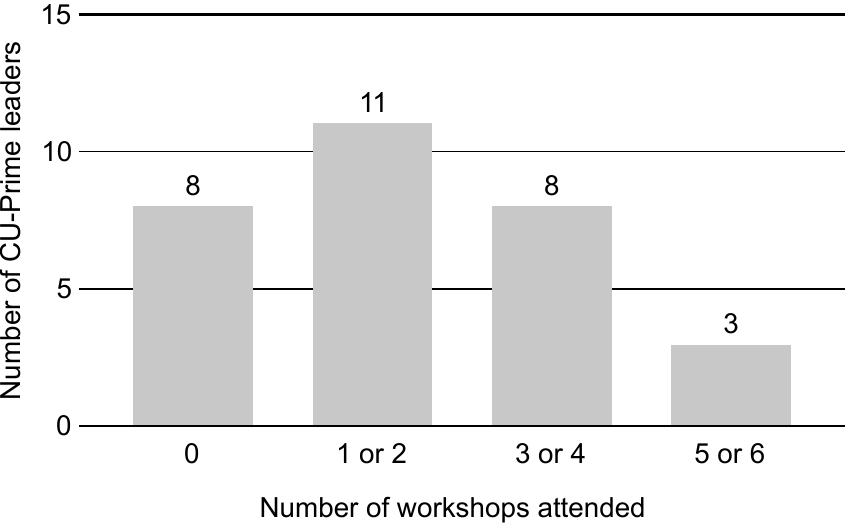}
	\caption{\label{fig:attendance}Workshop attendance. \drdf{Shown is the number of CU-Prime leaders (excluding workshop facilitators) who attended a particular number of workshops.}}
	\end{figure}
}

\newcommand{\tabA}{
	\begin{table}[t]
	\caption{\label{tab:topics}Workshop numbers, dates, titles, and attendance. Race and ethnicity were major themes in Workshops 1--3.}
	\begin{ruledtabular}
	\begin{tabular}{lllr}
	No. & Date & Topic & Att. \\ \hline
	1& 12/2014 & Women in physics: Statistics and biases & \drdf{8} \\
	2& 03/2015 & First-generation college students & 8 \\
	3& 06/2015 & Achievement gaps and the deficit model & 8 \\
	4& 09/2015 & Models of allyship and social change & \drdf{12} \\
	5& 12/2015 & Deconstructing the word ``diversity" & 6 \\
	6& 03/2016 & Summative workshop & \drdf{12} \\ \hline
	\multicolumn{3}{r}{Average:} & \drdf{9}
	\end{tabular}
	\end{ruledtabular}
	\end{table}
}

\begin{document}

\title{Learning to do diversity work: A model for continued education of program organizers}

\author{Dimitri R. Dounas-Frazer}
\affiliation{Department of Physics, University of Colorado Boulder, Boulder, CO 80309, USA}

\author{Simone A. Hyater-Adams}
\affiliation{ATLAS Institute, University of Colorado Boulder, Boulder, CO 80309, USA}

\author{Daniel L. Reinholz}
\affiliation{Department of Mathematics \& Statistics, San Diego State University, San Diego, CA 92182, USA}

\date{\today}

\maketitle


Physics and physics education in the United States suffer from severe (and, in some cases, worsening) underrepresentation of Black, \drdf{Latinx},\footnote{\drdf{We use the term ``Latinx" as a gender-inclusive alternative to the terms Latina and Latino.}} and \drdf{Native American} people of all genders and women of all races and ethnicities.\cite{NRC2013,AIP2012,NCES2011} \drdf{In this paper, we describe an an approach to facilitating physics students' collective and continued education about such underrepresentation; its connections to racism, sexism, and other dimensions of marginalization; and models of allyship that may bring about social change within physics.} Specifically, we focus on the efforts of undergraduate students, graduate students, and postdocs who are members of a student-run diversity-oriented organization in the \drdf{Physics Department at the University of Colorado Boulder (CU),} a large, selective, predominantly white, \drdf{public} university with high research activity. This group's education was accomplished through quarterly Diversity Workshops. Here we report on six Diversity Workshops that were co-designed and facilitated by the authors. We describe the context, motivation, and goals of the workshops, the theories underlying their design and implementation, and their content. In addition, we discuss workshop attendance and suggest strategies for maintaining high attendance in the future. Because the details of our workshops were tailored to the specific needs and interests of a particular student organization, our workshop agendas may not be widely applicable beyond our local context. 
\drdf{Nevertheless, our model, design principles, and facilitation strategies may be transferable to other contexts and provide inspiration to other diversity-oriented student groups}.

\section{Context, motivation, \& purpose}

\figA

The Diversity Workshops were one of several services offered by an organization \drdf{called \emph{CU-Prime}}. \drdf{CU-Prime} is a member of \emph{The Access Network (Access)},\footnote{The Access Network, http://accessnetwork.org} a network of seven university-based organizations characterized by student leadership and a commitment to improving diversity in the physical sciences through community building. \drdf{CU-Prime} is situated in \drdf{the CU Department of Physics, a large} department whose representation of undergraduate students who are female and/or who belong to underrepresented racial and ethnic groups is even lower than national averages: \drdf{between 2005 and 2014, 16\% of physics bachelor degree recipients at CU were women and 5\% were students from underrepresented minority groups. For reference, across the nation, women and underrepresented minorities comprise 20\% and 7\% of bachelor degree recipients, respectively, with fewer than 4\% of degrees awarded to women of color (including Asian American women).}\cite{AIP2012,NRC2013}

\drdf{CU-Prime} was founded in part as a response to the homogeneity of the physics department with respect to race, ethnicity, and gender. It was also founded to respond to a lack of community among undergraduate and graduate students of all genders, races, and ethnicities. While \drdf{CU-Prime} leadership originally consisted of a few physics graduate students, there are currently 33 \drdf{CU-Prime} leaders\drdf{, including all three authors of this paper}: 16 graduate students, 13 undergraduate students, and 4 postdoctoral researchers. About 50\% of leaders are white men and about 30\% are white women. About 14\% are men of color and 6\% are women of color. \drdf{The large leadership of CU-Prime is due, in part, to its inclusive leadership policy: anyone who attends an organizational meeting is able to fully participate in a leadership capacity.} Though some leadership roles within \drdf{CU-Prime} are paid through the university, most \drdf{CU-Prime} leaders do not receive any financial compensation for their work and all leaders volunteer with one or more aspects of the organization. \drdf{CU-Prime leaders do not have influence over departmental policy due to their role in CU-Prime. However, several CU-Prime leaders hold other leadership positions within the department, e.g., some are members of a committee dedicated to improving diversity in the CU Physics Department.}

Access member organizations offer multiple, mutually reinforcing services aimed at improving diversity and community among physics students: summer programs for incoming first-year undergraduate students;\cite{Dounas-Frazer2013TPT} academic-year coursework in which first- and second-year undergraduate students develop and carry out their own research projects;\cite{Gandhi2016} mentorship programs that pair graduate student mentors with undergraduate student mentees;\cite{Zaniewski2016} and seminar series where graduate students and professors present their research to a predominantly undergraduate audience. While the seminar series is open to all students regardless of demographic background, Access member organizations specifically recruit students from underrepresented groups into summer programs, coursework, and mentorship programs. \drdf{CU-Prime} runs a subset of these services.

The Diversity Workshops differ from the other services in an important way: the direct beneficiaries of the workshops are \drdf{CU-Prime} leaders themselves. These workshops provide \drdf{CU-Prime} leaders with the background and resources to make informed programmatic decisions about their organization. Courses in Gender Studies, Ethnic Studies, and so on are not typically part of the preparation of physics graduate students---including the leaders of \drdf{CU-Prime}---resulting in a need for additional education in the realm of equity in physics. The Diversity Workshops address this need by providing opportunities for \drdf{CU-Prime} leaders to (a) learn about the challenges and successes of students from underrepresented groups in physics, and to (b) apply this learning to the continued evolution of their organization.

\section{Design}

Our workshop model provided a dedicated space for \drdf{CU-Prime} leadership to discuss diversity-related topics while simultaneously respecting the fact that \drdf{CU-Prime} leaders are volunteers with many demands on their time. Based on conversations with \drdf{CU-Prime} leadership, we determined that quarterly workshops were appropriate for the organization's needs.  A flowchart for an idealized format of a workshop is presented in Fig.~\ref{fig:model}; in practice, workshops deviated slightly from this idealized model in idiosyncratic ways. According to this scheme, an individual workshop would last two hours and consist of five parts: pre-workshop homework, an Empathy Activity, a homework debrief, an Application Activity, and post-workshop feedback. A list of workshop topics is provided in Table~\ref{tab:topics}. In this section, we describe our rationale for the structure and content of the Diversity Workshops. To help clarify our ideas, we occasionally draw on examples from the workshop focused on the experiences of first-generation college students (Workshop~2 in Table~\ref{tab:topics}).

The design of the Diversity Workshops was informed by several complementary theories about the benefits and challenges of intergroup discussions. We define ``intergroup discussions" as discussions that satisfy two criteria: (i) the topic of discussion is one or more dimensions of marginalization; and (ii) the participants in the discussion include people who both do and do not belong to the corresponding marginalized group. For example, a discussion about racism among white people and people of color constitutes an intergroup discussion.

Because \drdf{CU-Prime} leadership is heterogeneous with respect to gender, race, and other identities, each Diversity Workshop was an intergroup discussion. Therefore, our designs incorporated the following six principles of intergroup discussions:
\begin{enumerate}
\item[P1.] Intergroup discussions can be unsafe for members of the marginalized group.\cite{Leonardo2010}
\item[P2.] Allies can develop through repeated exposure to counterideologies about marginalization.\cite{O'Brien2007} 
\item[P3.] Both scholarly work about marginalization and the lived experiences of students from the corresponding marginalized group are equally important.\cite{Ladson-Billings1995}
\item[P4.] Discussions of marginalization must focus on the intersections of race, gender, class, and other identities.\cite{Crenshaw1989}
\item[P5.] Empathy is a critical feature of productive intergroup discussions.\cite{Stephan1999}
\item[P6.] Discussions and resources must be tailored to the needs and goals of the organization.
\end{enumerate}
By ``unsafe," we mean that participation in intergroup discussions can carry emotional, professional, and even physical risks for people from the marginalized group.\cite{Leonardo2010} ``Allies" to a marginalized group are people who do not belong to the group but who work to minimize and/or eliminate the corresponding form of marginalization. We define ``counterideologies about marginalization" in contrast to prevailing ideologies that promote marginalization. For example, the idea that racial discrimination is rare is an ideology that upholds racism. This ideology underlies, for example, the belief that underrepresentation of Black students in post-secondary education is a purely socioeconomic phenomenon that has little or nothing to do with race.\cite{Bonilla-Silva2006} A related antiracist counterideology is characterized by the recognition that racial discrimination is both frequent and a major source of racial disparities in education and other contexts.\cite{O'Brien2007} ``Lived experiences" are the subjective experiences and personal narratives about marginalization told by people from the marginalized group.\cite{Ladson-Billings1995} Principles P1--P6 guide our intergroup discussions.

To minimize risk to members of marginalized groups participating in the Diversity Workshops (P1), participation was limited to \drdf{CU-Prime} leaders.\footnote{\drdf{Occasionally, students who were interested in joining CU-Prime leadership were invited to attend a Diversity Workshop. In total, 4 such students each attended 1 workshop.}} This ensured that workshop participants were people who already recognized lack of diversity in the physics department as a problem, were committed to improving diversity, and had established trusting and friendly relationships with one another through their volunteer work in \drdf{CU-Prime}. In addition, because \drdf{CU-Prime} leadership consists only of students and postdoctoral researchers, this restriction on participation effectively excluded tenured and tenure-track faculty members, non-tenure-track instructors, and departmental staff. Thus, the model we present here is not necessarily appropriate for use in settings where students, professors, and staff are co-learners. In these settings, additional precautions to minimize potential professional backlash would need to be employed. 

To ensure that \drdf{CU-Prime} leaders were repeatedly exposed to counterideologies about marginalization (P2), the Diversity Workshops were framed as an ongoing curriculum rather than a one-time intervention. Workshops were offered regularly four times per year: March, June, September, and December. In addition, successive workshops built off of one another, thus reinforcing their framing as a curriculum. For example, the content of a particular workshop was informed by feedback from previous workshops. Similarly, homework sometimes included reading the notes from previous workshops. 

Consistent with values P3 and P4, workshop themes, and hence homework content, focused on intersections of multiple identities. In addition, pre-workshop homework materials included peer-reviewed journal articles about a particular aspect of marginalization as well as blog posts and videos created by people from the corresponding marginalized groups. For example, to prepare for Workshop~2, participants were tasked with reading a journal article about retaining Latina/o and first-generation students in post-secondary education,\cite{Harrell2003} reading a blog post written by a first-generation student discussing the tension of having ``so many doors opened to you that four years just isn't enough time to truly indulge in them all,"\cite{Lattimore2014} and watching an audio slideshow in which first-generation students spoke about the challenges of balancing academic, family, and work-related responsibilities.\cite{WKCD2014}

\tabA

To improve the quality of intergroup discussions, we implemented Empathy Activities at the beginning of most workshops (P5). Empathy Activities involved playing a short video, either of a student speaking about their experience or of a spoken word poet speaking about a dimension of marginalization. After watching the video, participants were asked to silently and individually reflect on three questions: where is the speaker coming from, how do you think they feel, and how does that make you feel? Silent reflection was followed by small- and large-group discussions about participants' responses to the video. Empathy Activities set the stage for debriefing on homework, which typically took place during the middle of the workshop.

For example, in Workshop~2, participants watched a video blog in which a student named Manuel discussed his experiences as a first-generation college student: Manuel grew up in a Spanish-speaking community on the West Coast and ultimately left that community to attend college on the East Coast. Manuel compared his decision to leave home for college to his parents' decision to emigrate from Mexico in search of a better life. He acknowledged both the challenge of feeling separated from home and the strength of being bicultural, \emph{i.e.}, being able to bridge the culture of an Ivy League university and that of his hometown.\cite{Contreras2014} This video (and the subsequent discussion about participants' responses to the video) provided workshop participants with a specific story that they could relate to when discussing the statistics, strategies, and narratives presented in their homework materials.

As shown in Fig.~\ref{fig:model}, the final segment of each workshop involved an Application Activity during which participants applied what they'd learned to \drdf{CU-Prime} itself (P6). Application Activities took multiple forms. In Workshops~1 and 2, participants broke into three small groups: one focused on the \drdf{CU-Prime} seminar series, another on the course, and a third on the mentorship program. Each group discussed how to translate what they'd learned into practice. For example, during the Application Activity in Workshop 2, participants proposed creating opportunities to explicitly recognize and honor students' personal narratives: speakers in the seminar series could be asked to speak about their life-paths as well as their research projects, and the course could incorporate a panel discussion where upper-division undergraduate students from diverse backgrounds could talk about their transition to college life. In Workshops~3 and 4, the Application Activities involved discussions about the mission and identity of \drdf{CU-Prime} as a ``diversity organization." These activities reflect the content of the workshops, which focused on the benefits and pitfalls of various models for supporting diversity: people from marginalized and non-marginalized groups co-working towards change versus people from non-marginalized groups creating remedial spaces for those from marginalized groups. Finally, in Workshops~5 and 6, the Application Activities involved discussions of organizational values, success metrics, and accountability mechanisms.

While Application Activities created opportunities to tailor a particular workshop to the \drdf{CU-Prime} context, post-workshop feedback facilitated such tailoring on longer timescales. After each workshop, we solicited input from workshop participants about the content and structure of previous workshops as well as topics of interest for future workshops. This feedback mechanism helped us respond to the evolving needs and interests of \drdf{CU-Prime} leaders with respect to education about diversity-related issues.

\section{Facilitation}

We aligned our facilitation strategies with principle P1 in mind, taking care to prioritize the safety of students from marginalized groups. Here we describe four aspects of our facilitation: attending to participants' emotions, monitoring participants' language, incorporating multiple modes of communication, and furthering our own continued education.

Not only are intergroup discussions inherently risky for students from marginalized groups, but students from non-marginalized groups may also experience discomfort.\cite{DiAngelo2011} Moreover, because identity is multifaceted and our conversations focused on the intersections of multiple identities, an individual participant's identity may have both marginalized and dominant aspects simultaneously. In some cases, participants may need to temporarily excuse themselves from a workshop in order to process their emotional responses to an activity or discussion. Indeed, this happened in our workshops on multiple occasions. This can happen for a variety of reasons: a student of color may leave the room because someone said something racist and the facilitators neglected to respond appropriately, or a white student may leave the room because they need space to process a new realization about their own racial biases. In these cases, it can be useful to have a facilitator accompany the participant and check in with them. 

Since our discussions often touch on multiple identities at once---including invisible identities like status as a first-generation student---it is impossible to know why a particular individual is leaving the room without talking to them. In order for a facilitator to do so without disrupting the workshop, a team of facilitators is necessary. We found that a three-person team worked well for us: while one person facilitated a discussion or activity another could take notes, leaving a third facilitator to attend to any emotional issues that may arise.

Sometimes people say and do things that contribute to the marginalization of marginalized groups. Even the most well-meaning among us may do so despite our best intentions. Indeed, one or more \drdf{CU-Prime} leaders made unintentionally marginalizing comments in every Diversity Workshop. These comments were rarely (if ever) explicitly racist, sexist, or otherwise marginalizing; rather, they used race- and gender-neutral language. One example is the comment, ``I don't see race, I only see human beings." While this comment may seem innocuous, it is nevertheless an example of colorblind racism\cite{Bonilla-Silva2006} because it dismisses the reality of race and racism in the United States. As facilitators, we not only needed to be aware that this type of language exists, but also to actively listen for it during small- and large-group discussions. Whenever we heard comments that were implicitly racist, sexist, or otherwise marginalizing, we intervened by highlighting these unintended implications. The goal was to reduce the harm done to students from marginalized groups who were targeted by the comment.

Participation in intergroup discussions requires balance. For example, students from dominant groups should be mindful of how often they speak, neither speaking too often nor remaining silent.\cite{DiAngelo2012} As facilitators, one way to help people strike this balance is to be explicit about the importance of self-monitoring one's own participation. Another way to promote this balance is to incorporate multiple modes of communication in the workshops. In each workshop, we dedicated time to silent reflection and/or journaling, pairwise conversations, and whole group discussions. We also incorporated drawing and performance activities in Workshops 3 and 6, respectively. Our goal was to create multiple entry points for students to participate in the discussion as writers, speakers, drawers, performers, listeners, and observers.

Facilitating intergroup discussions can be challenging, and an experienced facilitation team is crucial to the success of these kinds of conversations. \drdf{We, the authors of this paper and facilitators of the Diversity Workshops, were mandatory reporters of harassment and had received anti-harassment training provided by the University of Colorado. In addition,} our facilitation team blended expertise in using performance as outreach to physics students (SAHA), facilitating conversations about diversity among physics professors, students, and department staff (DLR), and organizing student-run diversity-oriented groups  in the physical sciences (DRDF). Nevertheless, none of us is an expert in every form of marginalization or in every aspect of intergroup discussion. Therefore, during the three months between successive workshops, we met with one another to study, discuss, and learn about upcoming workshop topics and facilitation strategies. This commitment to ongoing self-education on the part of the facilitation team was imperative to the success of the Diversity Workshops.

\section{Attendance}

\figB
As a preliminary measure of whether we were successful in our goal of repeatedly exposing \drdf{CU-Prime} leaders to counterideologies about marginalization, we report attendance statistics for all six workshops. Attendance for each workshop is provided in Table~\ref{tab:topics} and a histogram describing attendance frequency is provided in Fig.~\ref{fig:attendance}. \drdf{In the table, figure, and discussion here, we exclude ourselves from attendance statistics since we facilitated the workshops and hence were not attendees. Accordingly, only 30 CU-Prime leaders could potentially attend the workshops.} Each workshop was attended by about \drdf{9 CU-Prime leaders}. In total, \drdf{22} unique \drdf{leaders} attended at least one workshop. Most leaders (\drdf{22 of 30}) participated in at least 1 workshop, and \drdf{about a} third of the leaders (11 of \drdf{30}) participated in at least 3 workshops. On average, about 60\% of workshop attendees provided us with feedback after each workshop. Every participant who provided feedback indicated that they would be interested in participating in future workshops. \drdf{This is consistent with the fact that about a third of the leaders opted to attend multiple workshops.}

The two workshops with the highest attendance were Workshops~4 and~6, as shown in Table~\ref{tab:topics}. The fourth workshop took place during an off-campus, multi-day retreat for \drdf{CU-Prime} leaders. Everyone who attended the retreat also participated in Workshop~4. The sixth workshop likely benefitted from its unique framing: whereas all other workshops were advertised as part of a sequence of workshops that built off of one another, Workshop~6 was advertised as an opportunity for newcomers to ``catch up" on material they missed. This framing was successful: seven \drdf{CU-Prime} leaders who had not previously attended a workshop participated in Workshop~6.

To ensure high attendance at future workshops, we will continue to offer a workshop at the retreat for \drdf{CU-Prime} leaders. In addition, we will frame each particular workshop both as a part of a broader curriculum \emph{and} as an acceptable entry point for newcomers to the curriculum.

\section{Summary}

We reported on the design of a sequence of Diversity Workshops for leaders of \drdf{CU-Prime}, a student-run group committed to improving race- and gender-based diversity in their physics department. We presented the model for our workshops, described six principles that informed our workshop design, and outlined four key aspects of our approach to facilitation. We hope that other student groups interested in learning more about diversity will adapt these principles and facilitation strategies to their own context.

\acknowledgments This material is based upon work supported by the Synberc Expanding Potential program under a Seed Project grant.

\bibliography{diversity_database}

\end{document}